\documentclass[
 reprint,
superscriptaddress,
 amsmath,amssymb,
 aps,
]{revtex4-2}

\usepackage{graphicx}
\usepackage{dcolumn}
\usepackage{bm}

\begin{document}

\preprint{APS/123-QED}

\title{``Beam \`a la carte": laser heater shaping for attosecond pulses in a multiplexed x-ray free-electron laser}



\author{Siqi Li}
\email{siqili@slac.stanford.edu}
\affiliation{SLAC National Accelerator Laboratory, Menlo Park, CA 94025, USA}
\author{Zhen Zhang}
\email{zzhang@slac.stanford.edu}
\affiliation{SLAC National Accelerator Laboratory, Menlo Park, CA 94025, USA}
\author{Shawn Alverson}
\affiliation{SLAC National Accelerator Laboratory, Menlo Park, CA 94025, USA}
\author{David Cesar}
\affiliation{SLAC National Accelerator Laboratory, Menlo Park, CA 94025, USA}
\author{Taran Driver}
\affiliation{SLAC National Accelerator Laboratory, Menlo Park, CA 94025, USA}
\affiliation{Stanford PULSE Institute, SLAC National Accelerator Laboratory, Menlo Park, CA 94025, USA}
\author{Paris Franz}
\affiliation{Applied Physics Department, Stanford University, Stanford, CA, 94305, USA}
\affiliation{SLAC National Accelerator Laboratory, Menlo Park, CA 94025, USA}
\affiliation{Stanford PULSE Institute, SLAC National Accelerator Laboratory, Menlo Park, CA 94025, USA} 
\author{Erik Isele}
\affiliation{Applied Physics Department, Stanford University, Stanford, CA, 94305, USA}
\affiliation{SLAC National Accelerator Laboratory, Menlo Park, CA 94025, USA}
\affiliation{Stanford PULSE Institute, SLAC National Accelerator Laboratory, Menlo Park, CA 94025, USA}
\author{Joseph P. Duris}
\affiliation{SLAC National Accelerator Laboratory, Menlo Park, CA 94025, USA}
\author{Kirk Larsen}
\affiliation{SLAC National Accelerator Laboratory, Menlo Park, CA 94025, USA}
\affiliation{Stanford PULSE Institute, SLAC National Accelerator Laboratory, Menlo Park, CA 94025, USA}
\author{Ming-Fu Lin}
\affiliation{SLAC National Accelerator Laboratory, Menlo Park, CA 94025, USA}
\author{Razib Obaid}
\affiliation{SLAC National Accelerator Laboratory, Menlo Park, CA 94025, USA}
\author{Jordan T O'Neal}
\affiliation{Physics Department, Stanford University, Stanford, CA, 94305, USA}
\affiliation{SLAC National Accelerator Laboratory, Menlo Park, CA 94025, USA}
\affiliation{Stanford PULSE Institute, SLAC National Accelerator Laboratory, Menlo Park, CA 94025, USA}
\author{River Robles}
\affiliation{Applied Physics Department, Stanford University, Stanford, CA, 94305, USA}
\affiliation{SLAC National Accelerator Laboratory, Menlo Park, CA 94025, USA}
\affiliation{Stanford PULSE Institute, SLAC National Accelerator Laboratory, Menlo Park, CA 94025, USA}
\author{Nick Sudar}
\affiliation{SLAC National Accelerator Laboratory, Menlo Park, CA 94025, USA}
\author{Zhaoheng Guo}
\affiliation{Applied Physics Department, Stanford University, Stanford, CA, 94305, USA}
\affiliation{SLAC National Accelerator Laboratory, Menlo Park, CA 94025, USA}
\affiliation{Stanford PULSE Institute, SLAC National Accelerator Laboratory, Menlo Park, CA 94025, USA} 
\author{Sharon Vetter}
\affiliation{SLAC National Accelerator Laboratory, Menlo Park, CA 94025, USA}
\author{Peter Walter}
\affiliation{SLAC National Accelerator Laboratory, Menlo Park, CA 94025, USA}
\author{Anna L. Wang}
\affiliation{Applied Physics Department, Stanford University, Stanford, CA, 94305, USA}
\affiliation{SLAC National Accelerator Laboratory, Menlo Park, CA 94025, USA}
\affiliation{Stanford PULSE Institute, SLAC National Accelerator Laboratory, Menlo Park, CA 94025, USA}
\author{Joseph Xu}
\affiliation{Argonne National Laboratory, Lemont, IL, USA}
\author{Sergio Carbajo}
\affiliation{SLAC National Accelerator Laboratory, Menlo Park, CA 94025, USA}
\affiliation{University of California at Los Angeles, Los Angeles, CA 90095, USA}
\author{James P. Cryan}
\email{jcryan@slac.stanford.edu}
\affiliation{SLAC National Accelerator Laboratory, Menlo Park, CA 94025, USA}
\affiliation{Stanford PULSE Institute, SLAC National Accelerator Laboratory, Menlo Park, CA 94025, USA}
\author{Agostino Marinelli}
\email{marinelli@slac.stanford.edu}
\affiliation{SLAC National Accelerator Laboratory, Menlo Park, CA 94025, USA}
\affiliation{Stanford PULSE Institute, SLAC National Accelerator Laboratory, Menlo Park, CA 94025, USA}

\date{\today}

\begin{abstract}
Electron beam shaping allows the control of the temporal properties of x-ray free-electron laser pulses from femtosecond to attosecond timescales. Here we demonstrate the use of a laser heater to shape electron bunches and enable the generation of attosecond x-ray pulses. We demonstrate that this method can be applied in a selective way, shaping a targeted subset of bunches while leaving the remaining bunches unchanged. This experiment enables the delivery of shaped x-ray pulses to multiple undulator beamlines, with pulse properties tailored to specialized scientific applications.
\end{abstract}

\maketitle

Free-electron lasers have emerged as the brightest sources of x-rays available \cite{emma2010first, ishikawa2012compact, prat2020compact, kang2017hard, decking2020mhz}, enabling the study of ultrafast phenomena with atomic resolution and femtosecond timing.
Two noteworthy advances in x-ray free-electron lasers (XFEL) technology include:  the generation of high power attosecond pulses~\cite{duris2020tunable, zhang2020experimental, marinelli2017experimental, prat2023coherent} and the demonstration of high repetition rate operation~\cite{decking2020mhz}. 
Attosecond x-ray free-electron laser represent a groundbreaking advancement in the field of ultrafast science, improving the peak brightness of attosecond light sources by more than six orders of magnitude~\cite{duris2020tunable}.
The advent of high-repetition rate XFELs, based on superconducting accelerator technology, represents a new significant leap forward, increasing the average power of XFEL sources by several orders of magnitude~\cite{decking2020mhz}. 
The potential to simultaneously support multiple users by efficiently distributing a high-repetition rate electron beam across numerous undulator lines will play an indispensable role in enabling sustained discovery science at XFEL facilities.
The extension of attosecond capabilities to this new generation of XFELs will unlock the full potential of attosecond x-ray science by massively increasing statistics and signal to noise ratio in nonlinear spectroscopy experiments. 

The generation of attosecond XFEL pulses involves manipulations of the electron beam phase space in order to generate a short high current spike, with a duration of$\sim 1$~fs and a peak current on the order of $10$~kA~\cite{duris2020tunable,zhang2020experimental}.
As we make progress in extending short-pulse capabilities to the next generation of XFELS, it is important to be able to deliver unique pulses ``on demand'' to a specific beamline while leaving the x-ray pulses delivered to other beamlines unaffected.
In this article we demonstrate such selective shaping of electron bunches for the generation of attosecond pulses. 
We show that by using a shaped laser pulse in the FEL laser heater we can selectively shape the bunches transported to the soft x-ray beamline of the LCLS-II free-electron laser, while leaving the operation of the hard x-ray beamline unperturbed. 
This is a key step towards the next generation of XFELs, with specialized undulator beamlines capable of delivering shaped pulses ``on demand".

Laser heaters are commonly used in XFELs to increase the slice energy spread of the electron beam in a controlled way. 
The increased energy spread suppresses collective instabilities that degrade the electron beam quality and decrease the performance of the XFEL \cite{saldin2004longitudinal}.
This is accomplished by the resonant interaction of the electrons with a laser pulse in a magnetic undulator placed in the middle of a magnetic chicane. \cite{huang2010measurements,ratner2015time}
The energy spread induced by the heater is proportional to the laser electric field, thus varying the temporal shape of the laser pulse can generate a time-dependent slice energy spread in the electron beam. 
Temporal shaping of the laser heater was shown to be an effective method for selective beam spoiling for the generation of femtosecond pulses, or for the generation of complex spectral modulations in seeded XFELs~\cite{marinelli2016optical, roussel2015multicolor}.
This temporal modulation of the slice energy spread can also be used to modulate the beam current profile. 
The perturbation of the current profile can be amplified by collective instabilities, resulting in one or more high-current spikes that can be used for the emission of attosecond pulses in an XFEL~\cite{cesar2021electron}. 
This method can also be used to generate stable short pulses in cavity-based XFELs \cite{tang2022electron}. 
An important feature of this laser-based shaping mechanism is that it can be arbitrarily timed with respect to the electron bunch. 
Therefore one can selectively shape a sub-set of the electron bunches produced by the linac, while leaving the other bunches unchanged. 
In an XFEL system with multiple beamlines, this selectivity can be used to generate a high-current spike in the bunches directed to a desired beamline, therefore delivering short pulses to a specific user while leaving the remaining XFEL  beamlines unperturbed.

\begin{figure*}[htb]
\centering
\includegraphics*[width=1.8\columnwidth]{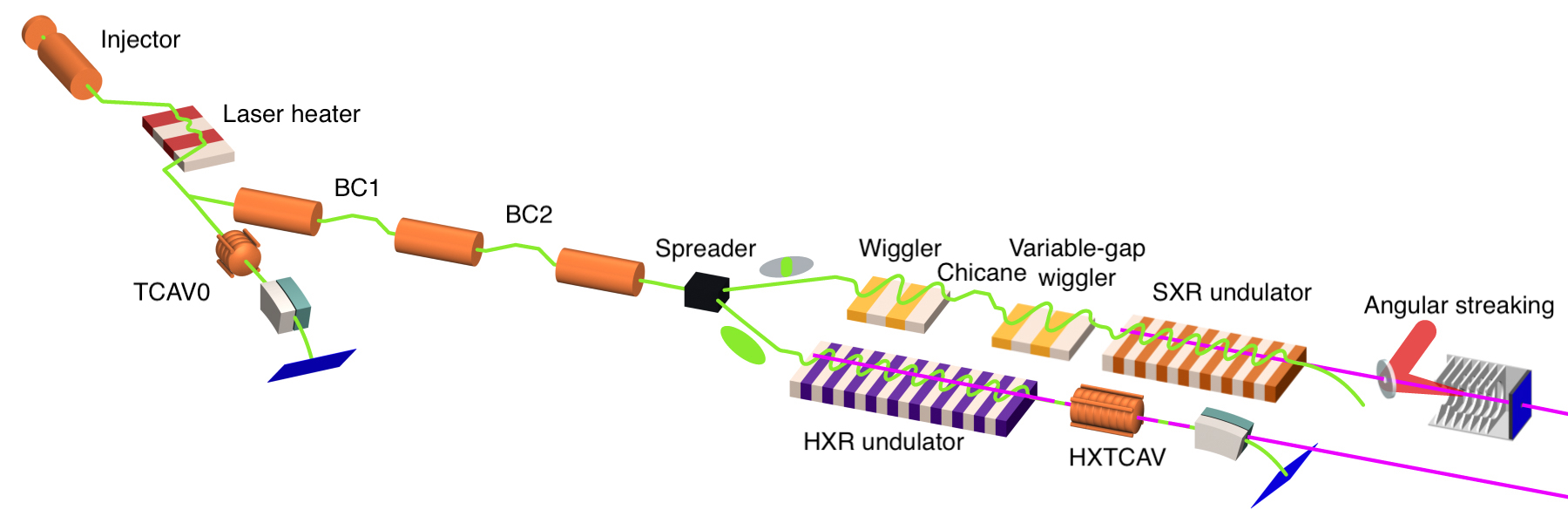}
\caption{\label{fig:scheme} Experimental layout of the laser heater shaping scheme at the LCLS-II.}
\end{figure*}

\begin{figure*}[htb]
\centering
\includegraphics*[width=1.8\columnwidth]{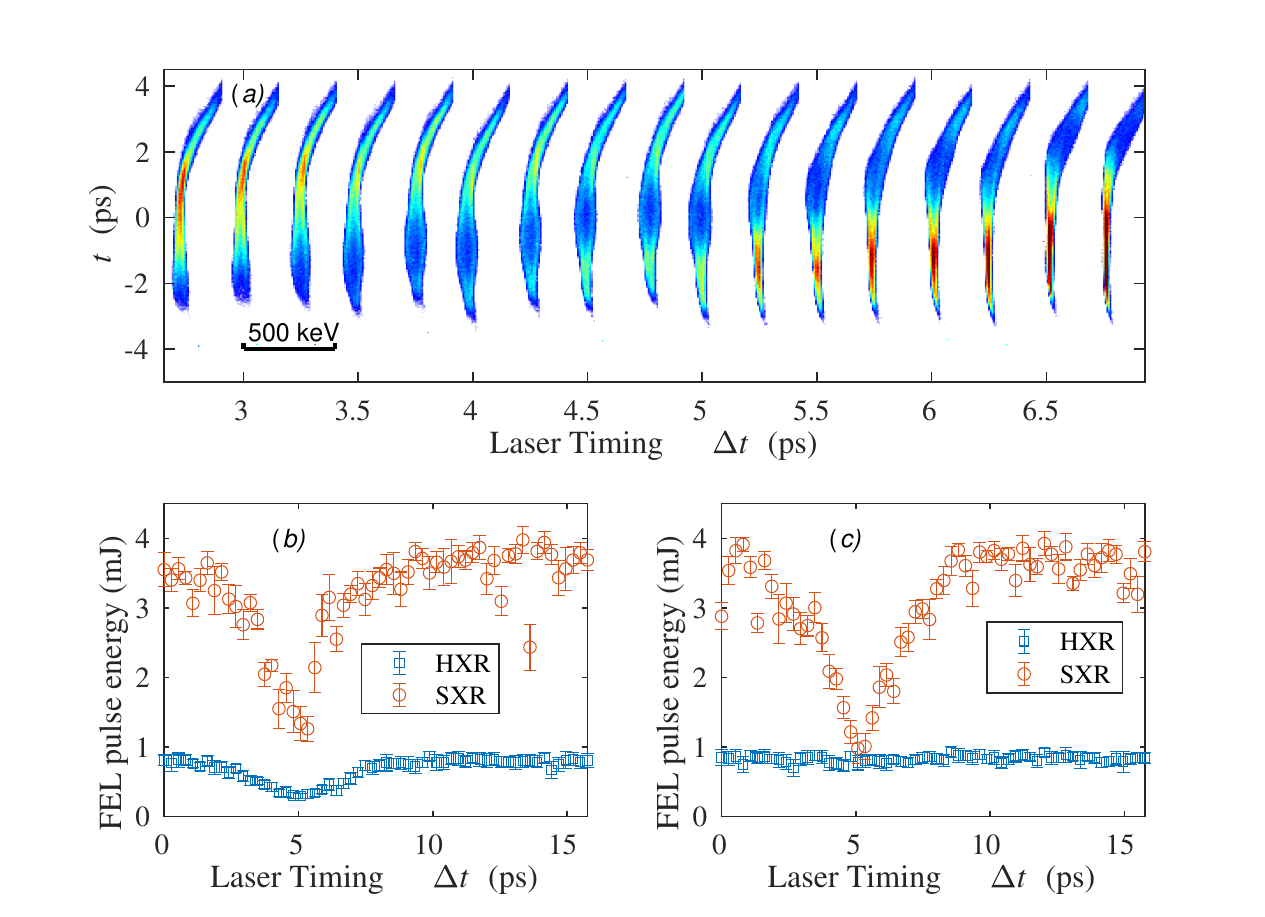}
\caption{\label{fig:tcav0} (a) Longitudinal phase space measured after injector by TCAV0 with different IR laser timing w.r.s. electron beam. Beam head lies to top. (b) Measurements of FEL lasing suppression at hard x-ray and soft x-ray undulators with laser heater shaping on all electron beams (120 Hz). (c) Measurements of FEL lasing suppression with selective laser heater shaping for soft x-ray line (the last 10 pulses for every 120 pulses per second).}
\end{figure*}

\begin{figure}[htb]
\centering
\includegraphics*[width=1\columnwidth]{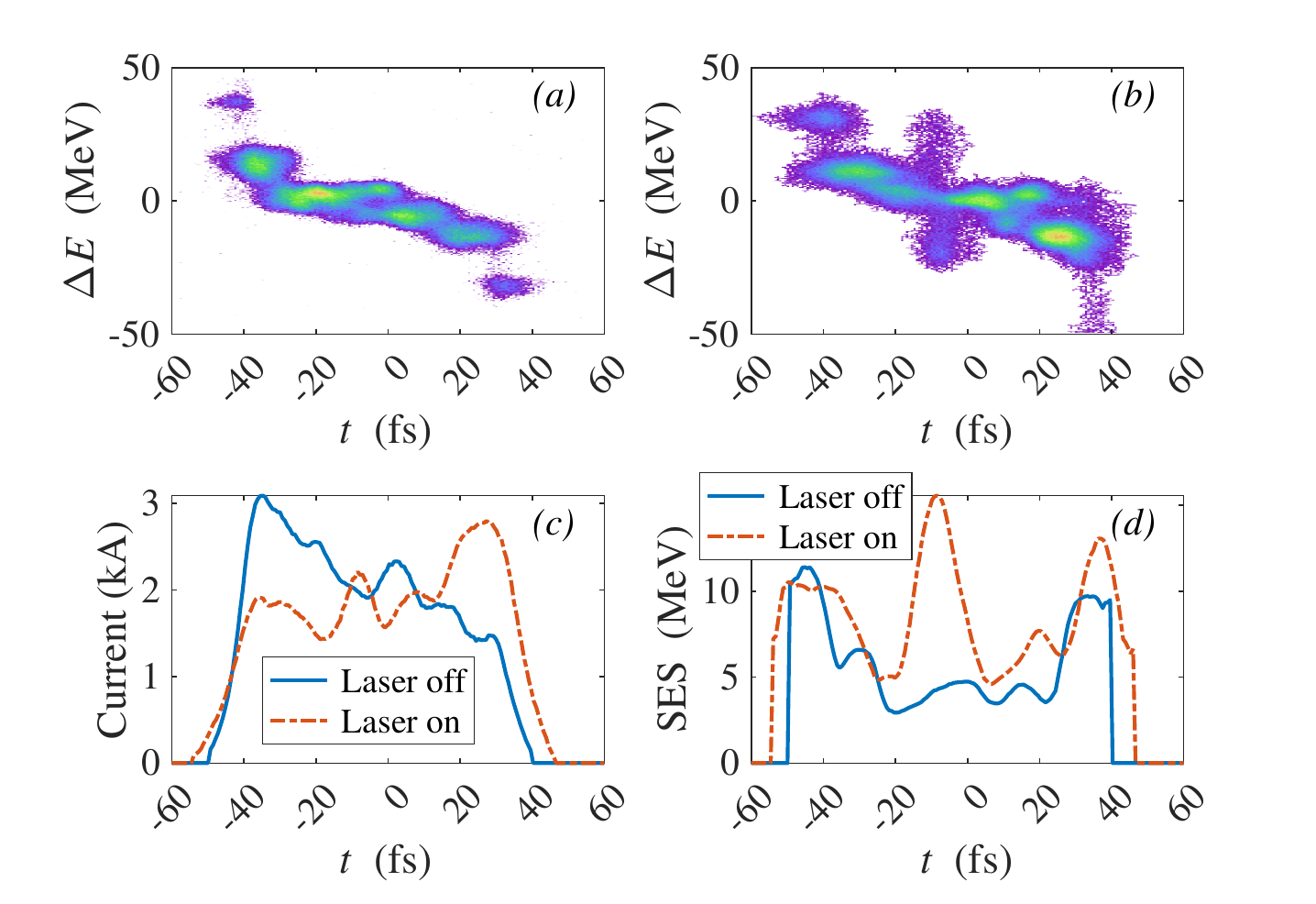}
\caption{\label{fig:streaking} Measurements of the longitudinal phase space of electron beam ($a$) w/o and ($b$) with later heater shaping. The beam current profiles ($c$) and slice energy spread (SES) profiles ($d$) are presented for comparisons.}
\end{figure}


\begin{figure}[htb]
\centering
\includegraphics*[width=1\columnwidth]{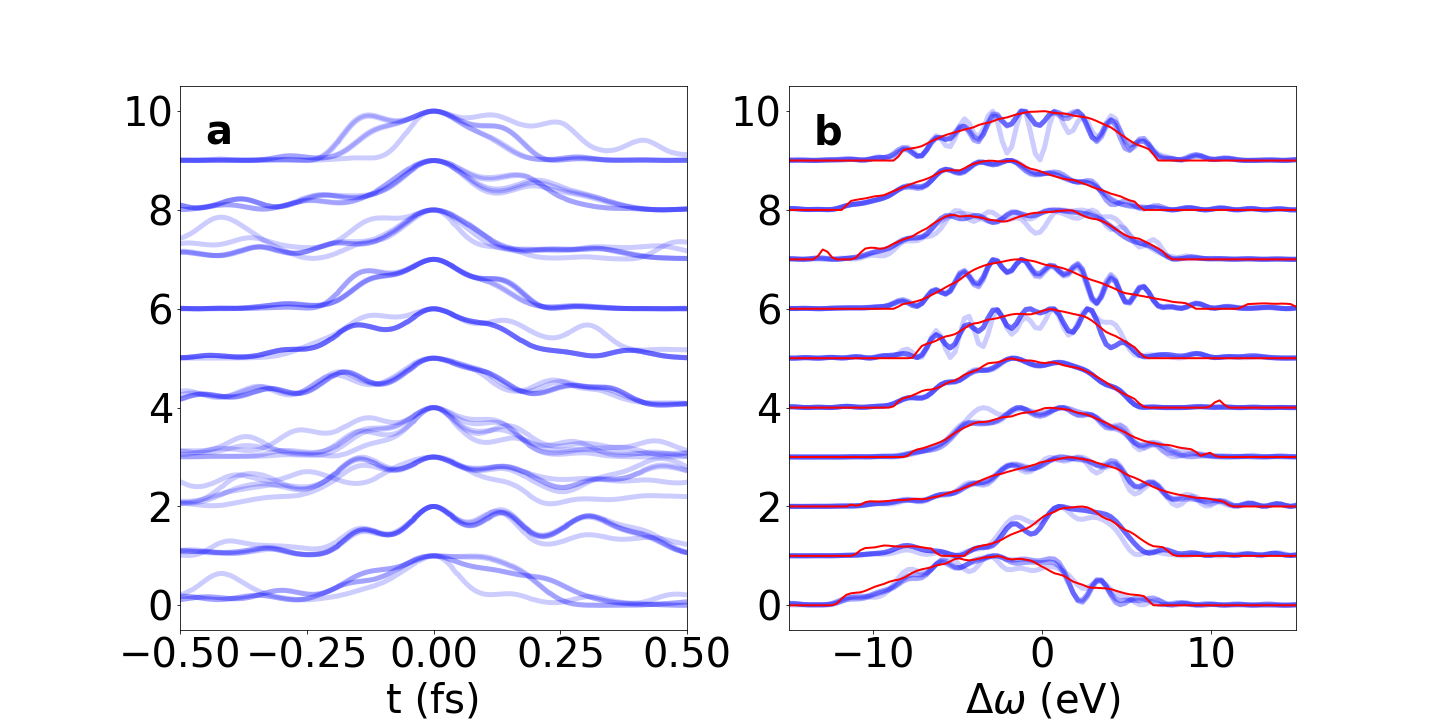}
\caption{\label{fig:streak_recon}Example pulse reconstruction using angular streaking. The left panel shows the single shot reconstructed pulse profile, and the right panel shows the single shot reconstructed spectrum, with the measured spectra in red. The shaded blue curve for each shot indicates the solution to the reconstruction algorithm with a different random seed.}
\end{figure}

\section{Experimental Results}
This experimental demonstration was conducted using the LCLS-II normal-conducting linac simultaneously feeding two undulator lines: the hard x-ray and the soft x-ray undulators. 
Two indepedent Ti:Sa laser systems are used in the laser heater, which we will term the heating laser and shaping laser.
The IR pulse from the heating laser is stretched to the duration of 20~ps to provide a uniform laser field over the electron bunch duration. The beam-shaping laser pulse is compressed to a shorter duration ($\sim$1.5 ps), and overlapped with the  pulse from the standard heating laser. The shaping pulse is shorter than the electron bunch and creates a short low-density region in the longitudinal phase-space. As the electron beam is accelerated and compressed, this shaped phase space seeds the microbunching instability, leading to the formation of a high-current spike \cite{cesar2021electron}.
It is worth emphasizing that because we use two independent lasers it is possible to achieve multiplexing by pulse picking. I.e. we can adjust the time-structure of the shaping laser so it only interacts with a sub-set of the electron bunches generated by the accelerator. 

 Figure\,\ref{fig:scheme} presents the schematic layout of the accelerator setup for this experiment. 
 The electron beam is generated in a high-brightness s-band photoinjector. The laser heater is placed at the end of the injector at an energy of approximately 135 MeV.
 The electron beam phase-space generated by the injector can be diagnosed using an s-band transverse deflecting cavity~(TCAV0) placed in a dedicated diagnostic beamline at the end of the LCLS photoinjector.
 After the laser heater, the electron beam is accelerated in the main linac and compressed in a two-stage compression system (the two bunch compressor systems are labeled as BC1 and BC2 in Fig.~\ref{fig:scheme}). Downstream of the linac, the electron beam can be sent to either undulator (SXR or HXR) at the desired beam repetition rate by using a fast kicker. The longitudinal phase-space is diagnosed with two X-band transverse deflecting cavities (XTCAV) placed at the end of both beamlines \cite{behrens2014few}. 
 We measure the attosecond x-ray pulses generated by the shaped bunches in the soft x-ray line using photo-electron angular streaking. 

The effect of the shaping laser on the longitudinal phase-space can be observed in Fig. \ref{fig:tcav0}, which shows the longitudinal phase-space measured after the photoinjector. We scan the relative timing of the shaping pulse and the electron bunch and observe a short overheated region in the beam phase-space moving from the head to the tail of the bunch. 
To demonstrate the ability to selectively shape the bunch profile, 
electrons bunches are sent to both the hard and soft x-ray undulators.
The electron bunches are produced at a repetition rate of $120$
~Hz, and the first 110 bunches are sent to the hard x-ray undulators with the rest sent to the soft x-ray undulators using a fast kicker.
Figures \,\ref{fig:tcav0} (b) and (c) show the emitted pulse energy as a function of shaping laser timing in both undulators. In the first data set, the shaping laser is operating at 120 Hz and interacting with all the bunches. In the second data set, the shaping laser interacts only with the last 10 electron bunches per second feeding the SXR undulator, while leaving the other 110 electron bunches untouched. When the shaping pulse is set to interact with all the electron bunches, the shaping mechanism affects the performance of both undulators. However, when the shaping pulse is only interacting with the bunches directed to the SXR line, the performance of the HXR line is not affected. We note that the shaping method reduces the overall pulse energy. This is because of the complexity of the resulting longitudinal phase space, in which different parts of the bunch have different time-energy correlation, requiring different undulator configurations for efficient lasing \cite{SaldinChirpTaper}.



Figure\,\ref{fig:streaking} shows the longitudinal phase-space measured at the end of the HXR undulator with the XTCAV. Panel (a) shows the phase-space with no shaping laser, while panel (b) shows the phase-space with shaping laser on and interacting with the HXR bunches.
The current and slice energy spread profiles with and without laser heater shaping are presented in Figs. 3 (c) and (d) respectively. A high current spike appears at the core part of the electron beam when the shaping laser is present. Note that the temporal resolution of the measurement ($\sim 1 fs$, root mean square) is insufficient to fully resolve the temporal shape of the spike, leading to an underestimate of the peak current. A distinct feature of the shaped electron beam phase space is the presence of a large time-energy correlation within the high-current spike induced by the longitudinal space charge force. This correlation is lost in the measurement due to the limited time resolution, and appears as a large uncorrelated energy spread in the center of the bunch.

\section{\label{attosecond xray}Attosecond x-ray FEL pulses}

\begin{table}
\caption{\label{tab:parameters} Beam and machine parameters.}
\begin{ruledtabular}
\begin{tabular}{lcr}
parameters & values (Fig~\ref{fig:streaking}) & (Fig.~\ref{fig:streak_recon})\\
\hline
charge (pC) & 165 & 180 \\ 
BC1 current (kA) & 0.195 & 0.195\\
BC2 current (kA) & 2 & 3.1 \\
electron beam energy (GeV) & 5 &6.5 \\
central photon energy (eV) & N/A  &715 - 725\\
\end{tabular}
\end{ruledtabular}
\end{table}

The high-current spike generated with heater shaping is employed for the generation of attosecond pulses in the SXR undulator, using the  methods described in \cite{cesar2021electron}. The details of the undulator and accelerator configuration are shown in the supplementary information. We operated the SXR undulator with shaped laser heater while delivering to the HXR line during user operation with the regular laser heater. Critically, the FEL performance of the HXR line was unaffected by the shaping experiment in the SXR line (See the Supplementary Information).
We observe single-spike x-ray pulses at a photon energy of 720 eV. The entire data set shows a median pulse energy of 24~$\mu$J. The spectrum of the x-ray pulses is measured by a variable line spectrometer (VLS)~\cite{hettrick1988resolving, obaid2018lcls} and demonstrates a median bandwidth of 7.5~eV full width half maximum (FWHM). 

We use an angular streaking diagnostic to measure the pulse profile and duration of the pulses on a single-shot basis~\cite{duris2020tunable, li2018characterizing, hartmann2018attosecond, zhao2022characterization}.
This measurement was conducted in the TMO hutch of the LCLS facility. The measurement and pulse reconstruction were performed using the methods described in \cite{duris2020tunable,li2018characterizing}.
We reconstruct the temporal shape of the x-ray pulse for a subset of the data based on the measured x-ray pulse energy. 
The shots used for reconstruction have a median pulse energy of 57~$\mu$J and the corresponding x-ray spectra have a median bandwidth of 9.3~eV FWHM.
~Figure~\ref{fig:streak_recon} shows the result of 10 individual pulse reconstructions, together with the measured spectrum. The  reconstructed full width half max pulse duration has a median of 368~as with a standard deviation of +/- 150~as. 
In this data set, the heating laser pulse energy was kept at 10~$\mu$J and the shaping laser pulse energy was kept at 46~$\mu$J. In a different data set where we continuously scanned the pulse energy of the shaping laser from 18~$\mu$J to 712~$\mu$J, the reconstruction consistently showed a full width half max pulse duration of 320~as with a standard deviation of 186~as. 

In conclusion we have demonstrated the generation of attosecond x-ray pulses using laser shaping of electron bunches. The pulses were characterized in the time domain with an angular streaking diagnostic. The median measured pulse duration was 368~as, with a median FWHM bandwdith of 9.3 eV. 
This method can be used to selectively shape a subset of electron bunches, with application to multiplexed XFEL facilities. Specifically, we applied laser heater shaping to the soft x-ray line of the LCLS-II XFEL while leaving the bunches used in the hard x-ray line unperturbed.
This method paves the way to shaped pulse generation on demand in high-repetition rate XFEL facilities serving several users at a time.

\begin{acknowledgments}
Use of the Linac Coherent Light Source (LCLS), SLAC National Accelerator Laboratory, is supported by the U.S. Department of Energy, Office of Science, Office of Basic Energy Sciences under Contract No. DE-AC02-76SF00515.
A.M., D.C., P.F., Z.G. and Z.Z. acknowledge support from the Accelerator and Detector Research Program of the Department of Energy, Basic Energy Sciences division. 
Z.G., R.R. and P.F. also acknowledge support from Robert Siemann Fellowship of Stanford University.
The effort from T.D.D., J.W., M.F.K, T.W., and J.P.C. is supported by DOE, BES, Chemical Sciences, Geosciences, and Biosciences Division (CSGB). 
J. X. acknowledges support from the Department of Energy, Office of Science, under contract number 
DE-AC02-06CH11357.

\end{acknowledgments}

\bibliography{main}

\end{document}


\preprint{AIP/123-QED}

\title{Supplementary material}





\maketitle


\section{\label{sec:level1}Pulse reconstruction}

\begin{figure}[htb]
\centering
\includegraphics*[width=0.7\columnwidth]{supp_cvmi (1).png}
\caption{\label{fig:supp_cvmi} An example cVMI image measurement and the reconstructed cVMI image following the reconstruction procedure described in~\cite{li2018characterizing,duris2020tunable}. }
\end{figure}

We follow the procedures described in \cite{li2018characterizing,duris2020tunable} for the cVMI reconstruction. We take a cVMI image for each shot, and we  select the shots based on pulse energy and streaking amplitude. A fitted and scaled background is removed from the original experimental measurement to account for low energy electrons from the x-ray and the IR streaking laser and electrons that hit the cVMI plates inside the detector chamber. The processed cVMI image (see panel a of Fig.~\ref{fig:supp_cvmi}) is then fed into our pulse reconstruction algorithm that produces a reconstructed profile of the attosecond x-ray pulse including its amplitude and phase, which we use to calculate the pulse duration presented in the main article.

\section{\label{sec:level1}Beamline configurations}
As depicted in Figure 1 of the main text, the electron beams, shaped within the laser heater, undergo a two-stage compression process before being directed to the SXR line. Prior to entering the SXR undulator for FEL lasing, the beams traverse a sequential arrangement, comprising a fixed-gap wiggler, a magnetic chicane, and a variable-gap wiggler. This configuration, as previously discussed by Zhang et al. \cite{zhang2020experimental}, constitutes a stage for current enhancement. The time delay of the chicane and the gap of the downstream wiggler were optimized to achieve a  higher current spike and a larger energy chirp. During the angular streaking experiment, the two fixed-gap wigglers before the chicane are moved into the beamline and the gap of the variable-gap wiggler is 25\,mm. The $R_{56}$ of the chicane is set to -0.87\,mm. The detailed parameters of the wigglers can be found in Table\,\ref{tab:parameters_wiggler}.

\begin{table}
\caption{\label{tab:parameters_wiggler} Parameters of fixed-gap and variable-gap wigglers.}
\begin{ruledtabular}
\begin{tabular}{lcr}
parameters & values  & unit\\
\hline
\multicolumn{3}{c}{fixed-gap wiggler} \\
\hline
wiggler period  & 55 & cm \\ 
number of periods & 12 & \\
wiggler strength $K_w$ & 46 & \\
\hline
\multicolumn{3}{c}{variable-gap wiggler} \\
\hline
wiggler period  & 35 & cm \\ 
number of periods & 6 & \\
wiggler gap & 25 & mm \\
wiggler strength $K_w$ & 24.7 & \\
\end{tabular}
\end{ruledtabular}
\end{table}

\begin{figure}[htb]
\centering
\includegraphics*[width=0.7\columnwidth]{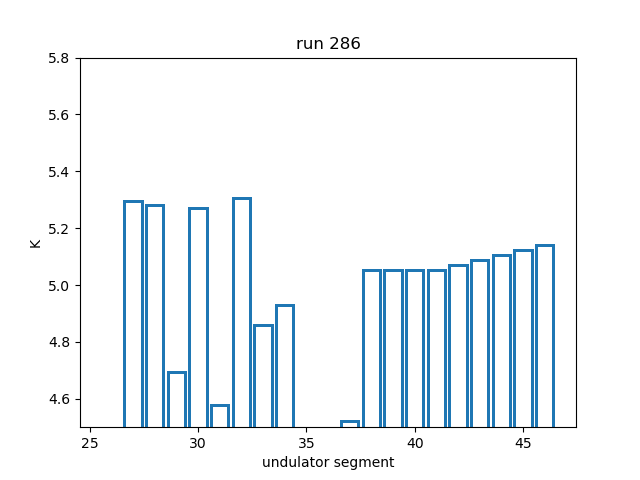}
\caption{\label{fig:supp_Kvsz} The undulator taper configuration during the measurement of attosecond x-ray pulses in the soft x-ray beamline. }
\end{figure}

The soft X-ray undulator taper is chosen to optimize the performance of the attosecond XFEL.
To achieve the shortest possible pulses we maximize the non-linear energy chirp induced by the undulator impedance \cite{baxevanis2018time}, following the method discussed in \cite{duris2020tunable}. 
To this end, we scramble the first several segments of the undulators so that no significant gain is accomplished in the first part of the undulator beamline, while exploiting the the undulator impedance to maximize the energy chirp.
We implement a flat taper followed by a linearly increasing taper towards the end of the undulator section, as shown in Fig.~\ref{fig:supp_Kvsz}. This undulator taper configuration optimizes the lasing and final power of the attosecond x-ray pulses.

\section{\label{sec:level1}Multiplexing operation}
\begin{figure}[htb]
\centering
\includegraphics*[width=1.0\columnwidth]{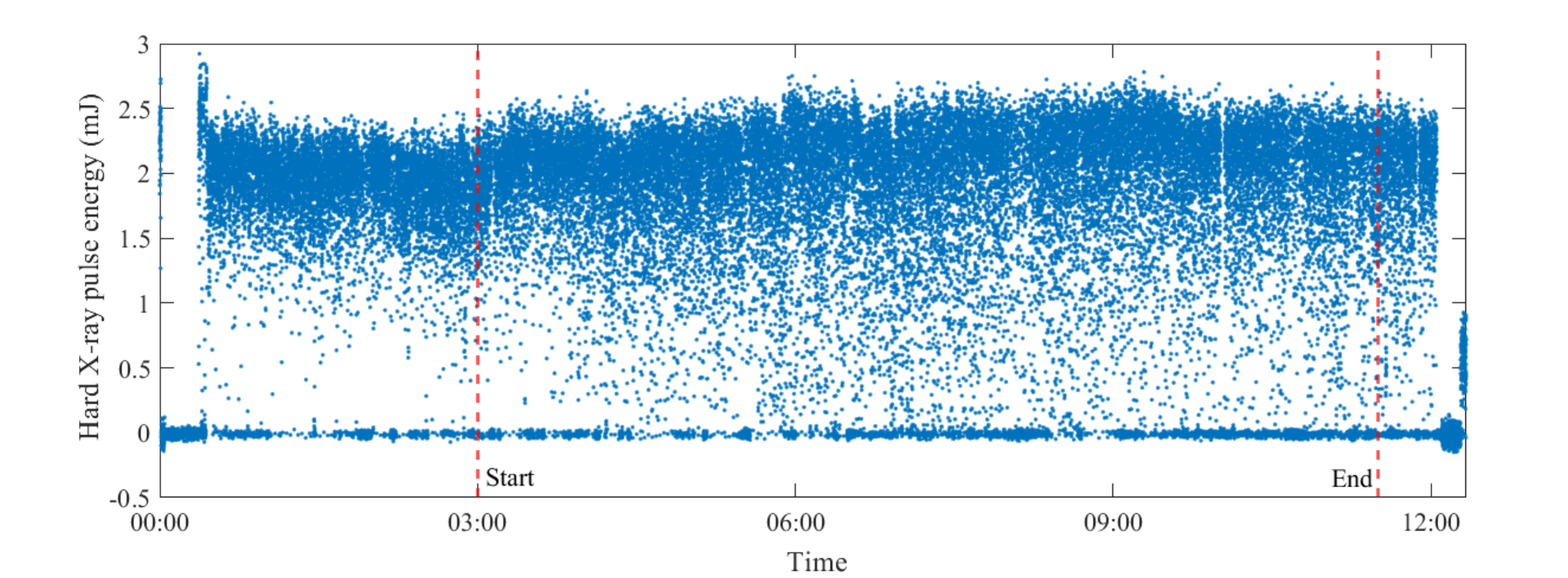}
\caption{\label{fig:supp_multiplexing} The hard X-ray pulse energy for user experiments during the laser heater shaping experiments at the soft X-ray line. The red lines in the figure represent the start and end of the laser heater shaping experiment, respectively. }
\end{figure}

The laser heater experiment to take angular streaking data was carried out with the 10-Hz electron beams directed towards the SXR line, while the additional 110-Hz beams were sent to HXR line for user experiments. Figure\,\ref{fig:supp_multiplexing} illustrates the HXR pulse energy during the commissioning of the laser heater shaping in the SXR line. This demonstrates the compatibility of the laser heater shaping with multiple undulator lines, showcasing its effectiveness in a multiplexed operational setup.

\nocite{*}
\bibliography{main_bib}